\documentclass[pra,a4paper,twocolumn,showpacs,superscriptaddress]{revtex4}
\usepackage{amsmath}
\usepackage{amsfonts}
\usepackage{graphicx}
\usepackage{longtable}
\begin{document}

\title{Kraus representation of quantum evolution and fidelity \\ as manifestations of Markovian and non-Markovian avataras} 
\author{A. K. Rajagopal} 
\affiliation{Inspire Institute Inc., Alexandria, Virginia, 22303, USA.}
\author{A. R. Usha Devi}
\email{arutth@rediffmail.com}
\affiliation{Department of Physics, Bangalore University, 
Bangalore-560 056, India}
\affiliation{Inspire Institute Inc., Alexandria, Virginia, 22303, USA.}
\author{R. W. Rendell} 
\affiliation{Inspire Institute Inc., Alexandria, Virginia, 22303, USA.}
\date{\today}

\begin{abstract} 
It is shown that the fidelity of the dynamically evolved system  with its earlier time density matrix provides a signature of non-Markovian dynamics. Also, the fidelity associated with the initial state and the dynamically evolved state is shown to be larger in the non-Markovian evolution compared to that  in the corresponding Markovian case. Starting from the Kraus representation of quantum evolution, the Markovian and non-Markovian features are discerned in its short time structure. These two features are in concordance with each other and they are illustrated with the help of four models of interaction of the system with its environment. 
\end{abstract}
\maketitle
\section{Introduction} 
The central theme of open quantum systems and their dynamical properties is to develop a description of the interaction of a quantum system with its environment~\cite{Breuer}. The significance of this area of research has been known for a long time and need not be emphasized, as indicated by a large body of literature on the subject. Most recently this has been a subject of intense study. A general form of the local time master equation describing this is given by Chruscinski and Kossakowski~\cite{CK}, which also gives references to the literature. They define clearly the meaning of the terms "Markovian" and "non-Markovian" incarnations of evolution as the absence and presence respectively of the initial time in the local generator of the master equation. Several manifestations of non-Markovianity have been proposed recently~\cite{Cirac,B2,Angel}, where non-Markovian reflections  are  recognized based on the departure of the evolution from strict Markovianity. While an abstract framework to identify 
whether a given quantum dynamical channel is Markovian or not has been put forth in Ref.~\cite{Cirac},  recently Breuer et. al.~\cite{B2} proposed  a quantification based on the maximum increase of the distinguishability  of two different initial quantum states  over the entire dynamical evolution. Evaluation of this measure, however, requires optimization of the total increase of the trace distance over all pairs of initial states. More recently~\cite{Angel}, deviations from Markovianity, in terms of the specific dynamical behavior of quantum correlations -- when part of an entangled system evolves under  a trace preserving completely positive quantum channel -- has been explored.  When complete tomographic information about the dynamical map is available a necessary and sufficient condition of non-Markovianity is also formulated~\cite{Angel}. All the above manifestations of non-Markovianity are built mainly by identifying deviations from the characteristic property of a Markovian channel --  being an element of one-parameter continuous, memoryless, completely positive semigroup. In this paper, we propose {\em fidelity difference} as a  non-Markovian incarnation -- which is yet another significant feature capturing the departure from the Markovian semigroup property of evolution.

We begin with the well-known Kraus representation~\cite{Kraus} of the reduced density matrix of the system interacting with an environment. In Sec.~II we begin by recalling the known result~\cite{preskill,RS} that if the Kraus operators exhibit a small-time dependence of a particular form, the Markovian master equation -- known as the Lindblad-Gorini-Kossakowski-Sudarshan  (LGKS)~\cite{Lindblad,GKS} master equation --  is recovered. This observation points towards the generality of the Kraus representation of the quantum evolution in subsuming both the Markov and Non-Markov versions depending on the structure of interaction between the system and its environment. 

In Sec.~III,  we propose  to use fidelity~\cite{Fidelity} $F[\rho(t), \rho(t+\tau)]$ as a measure to examine the nature of propensity of  the evolved density matrix $\rho(t+\tau)$ with the earlier time  density matrix $\rho(t)$. This offers a direct approach to the conventional view of Markovianity, namely that the fidelity $F[\rho(t), \rho(t+\tau)]$ would increase from its initial value $F[\rho(0), \rho(\tau)]$ and approach unity asymptotically. This is thus a   test,  for any deviation from this behavior would reflect non-Markovian incarnation. We also bring out the significance of our fidelity test  in comparison with  the recently proposed trace distance based quantification of non-Markovianity~\cite{B2}.    

In Sec.~IV, we illustrate our results through some examples. Here, we have considered the  exactly known models of Kraus representation given by Yu and Eberly, who investigated the issue of  sudden death of entanglement in the two-qubit system, evolving under  Markovian~\cite{YE1} and non-Markovian~\cite{YE2} environments. The small time nature of these model  Kraus representations illustrate the Markovian and non-Markovian natures in these examples. The fidelity $F[\rho(t),\rho(t+\tau)]$ in the Markovian case is shown to increase with time $t$, as expected. On the other hand, in the non-Markovian limit, we show that the fidelity difference $F[\rho(t),\rho(t+\tau)]-F[\rho(0),\rho(\tau)]$  fluctuates between positive and negative values -- bringing out the essence of non-Markovianity.  Further, we observe that the fidelity $F[\rho(0), \rho(t)]$ -- which corresponds to the memory of the initial state  carried by the dynamically evolving state  - is  larger in the non-Markovian limit, when compared with that in the Markovian case in this model.    

We also investigate another exactly known Kraus representation~\cite{AKRR, AKRU} of the Jaynes-Cummings  model of interaction of a qubit with the radiation field.  Unlike the other dynamical models discussed here, this example is exactly solvable and starts with the full Hamiltonian for which the unitary evolution operator can be constructed and as such, we examine here the corresponding dynamical equations associated with both the atom and photon systems. We explore the small time behavior of the Kraus operators in this model to recognize the Markovianity and secondly, we identify that the fidelity difference $F[\rho(t), \rho(t+\tau)]-F[\rho(0),\rho(\tau)]$ of the atom, initially in an excited state,  fluctuates between positive and negative values during evolution -- which is a clear signature of non-Markovianity. 

Recently, Chruscinski and Kossakowski (CK)~\cite{CK2} considered an interesting illustration  of non-Markovian dynamics of a single qubit, either  through a non-local master equation with a memory kernel  or equivalently via a seemingly simpler local in time equation. Both the descriptions are complimentary to each other: while the non-local equation involves a time-independent memory kernel, the corresponding local approach is governed by a highly singular generator. In other words, it has been illustrated that non-Markovianity manifests differently in local and non-local approaches. Here we show that in this example too the fidelity diference function captures the essense of non-Markovianity.            

Our analysis of these examples bring out typical characteristics of non-Markovian dynamics. Sec.~V is devoted to concluding remarks.  

\section{Kraus representation of quantum dynamics}
For simplicity of presentation, we  drop the system and environment Hamiltonians and consider only their mutual interaction. 
The dynamics of a system density matrix interacting with an environment is given in terms of Kraus representation as 
\begin{equation} 
\label{kraus}
\rho(t)=\sum_i K_i(t)\rho(0) K_i^\dag(t),
\end{equation}  
with the unit trace condition ${\rm Tr}[\rho(t)]=1$ leading to 
\begin{equation}
\label{sum}
\sum_i\, K_i^\dag K_i=I, 
\end{equation}  
$I$ denoting the identity matrix. 

We first recall~\cite{preskill, RS}, how the well-known LGKS master equation describing Markovian dynamics is obtained from Eq.~(\ref{kraus}). Following Preskill~\cite{preskill}, we separate one term, say $K_0(t)$ in the sum over $i$ in Eq.~(\ref{kraus}), and choose the rest of the terms $K_i(t), \ i\neq 0$ to have the following forms for small time $t$: 
\begin{equation}
\label{smtime}
K_i(t)\approx \sqrt{t}\, L_i, \ \ i\neq 0,  
\end{equation} 
when Eq.~(\ref{sum}) reduces to 
\begin{equation}
\label{k0} 
K_0(t)\approx I-\frac{t}{2}\, \sum_{i\neq 0} L_i^\dag L_i. 
\end{equation} 
Expressing the Kraus operators in the short time limit  in terms of the new $L$-operators, (\ref{kraus})  takes the standard LGKS form, termed as the Markovian master equation: 
\begin{eqnarray}
 \label{Lindblad} 
 \rho(t)&-&\rho(0)\approx t\, {\cal L}_M\, \rho(0), \nonumber \\ 
 {\rm i.e.,}\  \ \frac{d\rho}{dt}&=&{\cal L}_M\rho=\sum_{i\neq 0} \left[L_i\rho L^\dag_i-\frac{1}{2}\, (L_i^\dag L_i\rho+\rho\, L_i^\dag L_i)\right]. \nonumber \\
\end{eqnarray} 
(Derivation of the master equation (\ref{Lindblad})  from the Kraus representation (\ref{kraus}) in similar lines as above is also outlined in Ref.~\cite{RS}).  

It may be pointed out that in Ref.~\cite{CK2}, a
complete phenomenological treatment of local time evolution of open
quantum systems, based on a generalization of LGKS representation of
Markovian dynamics is discussed. This basically entails a local time-dependent
prefactor in the RHS of Eq.~(\ref{Lindblad}). A generalized non-Markovian master equation, which is local in time, has also been derived in Ref.~\cite{RS} and the short-time memory effects, retained from the
environment, are shown to lead to dissipations deviating from typical Markovian features~\cite{RSnote}.       
In the subsequent discussions in Sec.~IV, we explore if the Kraus operators exhibit the desired small time behavior leading to LGKS master equation or not in four different examples. 

\section{Fidelity and its implication for Markovianity} 

Following Jozsa~\cite{Fidelity}, we define the fidelity $F[\rho(t), \rho(t+\tau)]$ as the propensity of finding the state $\rho(t)$ in the later time state  
$\rho(t+\tau),\ \tau>0$: 
\begin{equation}
\label{fidelity} 
F[\rho(t),\rho(t+\tau)]=\left\{{\rm Tr}\left[\sqrt{\sqrt{\rho(t)}\rho(t+\tau)\sqrt{\rho(t)}}\right]\right\}^2, 
\end{equation}  
which is bounded by $0\leq F[\rho(t),\rho(t+\tau)]\leq 1$ and satisfies the symmetry property, $F[\rho(t),\rho(t+\tau)]=F[\rho(t+\tau),\rho(t)]$. 

Fidelity obeys another significant property i.e., monotonicity~\cite{NC}: 
\begin{equation}
F(\Lambda\,\rho_1,\Lambda\,\rho_2)\geq F(\rho_1,\rho_2)  
\end{equation} 
where $\Lambda$ denotes a completely positive map -- which serves as a characteristic feature of Markovian dynamics as indicated below.   

Recall that the Markovian evolution guarantees a completely positive, trace preserving dynamical map $\Lambda(t)$, 
\begin{equation}
\rho(0) \rightarrow \rho(t)=\Lambda(t)\rho(0), 
\end{equation}   
which also forms a one parameter semigroup obeying the composition law~\cite{CK, Cirac, B2, Angel, RS, CK2} 
\begin{equation}
\label{comp}
\Lambda(t_1)\Lambda(t_2)=\Lambda(t_1+t_2), \ t_1,t_2\geq 0,
\end{equation} 
a characteristic feature of Markovian dynamics. 
Therefore, it is clear that the fidelity function  $F[\rho(t),\rho(t+\tau)]$ involving the system density matrix evolving under Markovian dynamics satifies the inequality~\cite{note}, 
\begin{eqnarray}
\label{inequality}
F[\rho(t),\rho(t+\tau)]\equiv F[\Lambda(t)\rho(0),\Lambda(t)\rho(\tau)] \nonumber \\ 
\Rightarrow F[\rho(t),\rho(t+\tau)]\geq F[\rho(0),\rho(\tau)].
\end{eqnarray}
Any violation of this inequality is a clear signature of non-Markovian dynamics -- indicating that the associated dynamical map {\em does not} obey the the composition law (\ref{comp}) -- and hence, the dynamics has inbuilt memory effects. Deviation from the trend (\ref{inequality}) is, however, a sufficient -- though not necessary -- reflection of non-Markovianity. We propose to examine non-Markovianity in terms of the {\em fidelity difference} function 
\begin{equation}
\label{fd} 
G(t, \tau)= \frac{F[\rho(t),\rho(t+\tau)] - F[\rho(0),\rho(\tau)]}{F[\rho(0),\rho(\tau)]},
\end{equation} 
negative values of which  necessarily imply non-Markovianity.   We  identify the non-Markovian signature in terms of the fidelity difference function in some dynamical models  in Sec.~IV. 

It is worth pointing out the distinction between the quantification proposed by Breuer et. al.~\cite{B2} and our fidelity test of non-Markovianity proposed here. It is  well-known that the distinguishability of two states $\rho_{1},\ \rho_2$, measured in terms of the trace-distance $D(\rho_1,\rho_2)=\frac{1}{2}\vert\vert \rho_1-\rho_2\vert\vert$  never increases~\cite{NC} under all completely positive, trace preserving maps i.e., 
$D(\Lambda\,\rho_1,\Lambda\,\rho_2)\leq D(\rho_1,\rho_2)$. If the pair of states $\rho_{1,2}$ are evolving under the influence of a dynamical Markovian map i.e., $\rho_{1,2}(t)\equiv \Lambda(t)\rho_{1,2}(0)$,  the semi-group composition law (\ref{comp})  imposes that~\cite{B2, note2} 
$D[\rho_1(t+\tau),\rho_2(t+\tau)]\equiv D[\Lambda(\tau)\rho_1(t),\Lambda(\tau)\rho_2(t)]
 \leq D[\rho_1(t),\rho_2(t)],$ 
for all $t,\tau\geq 0$. This decisive property of Markovian processes, viz., {\em the trace distance of  any fixed pair of quantum states never increases}, has been employed in Ref.~\cite{B2} to uncover the non-Markovian feature in open system dynamics, in terms the following quantity 
\begin{eqnarray*}
{\cal N}&=& {{\rm max}\atop {\rho_{1,2}(0)}}  \int_{\sigma>0}\, dt\, \sigma[t,\rho_{1,2}(0)],\\ \sigma[t,\rho_{1,2}(0)]&=&\frac{d}{dt}\,D[\rho_1(t),\rho_2(t)],
\end{eqnarray*}
which measures the total increase of the trace-distance between any optimal pair of states $\rho_{1,2}$ during the entire time evolution. Evidently, this quantification requires optimization over the set of all initial {\em pairs} of states $\rho_{(1,2)}(0)$. Here, we have exploited the divisibility property (\ref{comp}) of the Markovian map differently, as illustrated in (\ref{inequality}), to obtain an inequality for the overlap $F[\rho(t),\rho(t+\tau)]$,   involving a dynamically evolving quantum state $\rho(t+\tau)$ and {\em its earlier time version} $\rho(t)$ -- in contrast to that for the trace distance of {\em  pairs} of  states $\rho_1, \rho_2$ under time evolution as in Ref.~\cite{B2}.

\section{Dynamical models}

We now proceed to explicitly investigate the small time behavior of Kraus representations and the nature of the fidelity functions in some simple models of open system dynamics. 
\medskip 

\noindent (a) Yu and Eberly~\cite{YE1} considered the following model Kraus operators for a simplified dynamical system of  two qubits interacting with environment along with its initial density matrix, for which we can verify both the small time limit to see if we can get LGKS Master equation and also fidelity to assure us of the interpretation of Markovian evolution or otherwise in our view. The initial state of the two qubit system is chosen to be in the simple form
 \begin{equation}
\rho_{AB}=\frac{1}{9}\, \left(\begin{array}{cccc} 
1 & 0 & 0 &0 \\ 
0 & 4 & \lambda & 0 \\ 
0 & \lambda & 4 & 0 \\ 
0 & 0 & 0 & 0 
\end{array} \right) 
\end{equation}
where $0\leq \lambda\leq 4$. The Kraus operators corresponding to the dynamical evolution of the qubits are given by, 
\begin{eqnarray}
\label{kr0}
K_0(t)&=&\left(\begin{array}{cccc} 
\gamma^2(t) & 0 & 0 &0 \\ 
0 & \gamma(t) & 0 & 0 \\ 
0 & 0 & \gamma(t) & 0 \\ 
0 & 0 & 0 & 1 
\end{array} \right),\nonumber \\    
K_1(t)&=&\left(\begin{array}{cccc} 
\gamma(t)\omega(t) & 0 & 0 &0 \\ 
0 & 0 & 0 & 0 \\ 
0 & 0 & \omega(t) & 0 \\ 
0 & 0 & 0 & 0
\end{array} \right), \nonumber \\ 
K_2(t)&=&\left(\begin{array}{cccc} 
\gamma(t)\omega(t) & 0 & 0 &0 \\ 
0 & \omega(t) & 0 & 0 \\ 
0 & 0 & 0 & 0 \\ 
0 & 0 & 0 & 0 
\end{array} \right), \nonumber \\
   K_3(t)&=&\left(\begin{array}{cccc} 
\omega^2(t) & 0 & 0 &0 \\ 
0 & 0 & 0 & 0 \\ 
0 & 0 & 0 & 0 \\ 
0 & 0 & 0 & 0
\end{array} \right),
\end{eqnarray}   
where $\gamma(t)={\rm exp}[-\Gamma t/2],\ \omega(t)=\sqrt{[1-\gamma^2(t)]};$ $\Gamma$ represents the strength of the environmental transverse noise.    
The dynamically evolved two qubit density matrix is given by,
\begin{eqnarray}
\label{ye1rho}
\rho_{AB}(t)&=&\sum_{i=0}^{3}K_i(t) \rho(0) K_i^\dag(t) \nonumber \\
&=&\frac{1}{9}\, \left(\begin{array}{cccc} 
1 & 0 & 0 &0 \\ 
0 & 4 & \lambda\, \gamma^2(t) & 0 \\ 
0 & \lambda\,\gamma^2(t) & 4 & 0 \\ 
0 & 0 & 0 & 0 
\end{array} \right). 
\end{eqnarray}
\begin{figure}
\includegraphics*[width=3in,keepaspectratio]{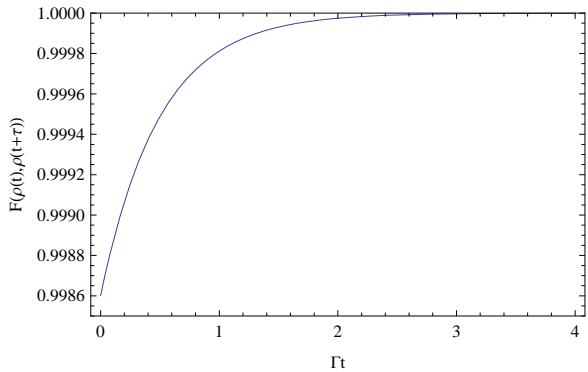}
\caption{(Color online). Fidelity $F[\rho(t),\rho(t+\tau)]$ of the dynamical state $\rho(t+\tau)$ with its earlier time density
matrix $\rho(t)$, as a function of dimensionless scaled time $\Gamma t$. Here, we have chosen $\Gamma\tau=1,\ {\rm and}\ \lambda=0.5$. The fidelity increases from its initial value $F[\rho(0),\rho(\tau)]$ and approaches 1 in the limit $\Gamma t\rightarrow \infty$ -- as anticipated in Markovian dynamics. All quantities are dimensionless.}
\end{figure}

In the small time limit i.e., $\Gamma t<<1$, we have, 
$\omega(t)\approx\sqrt{\Gamma\, t},\ \gamma(t)\approx (1-\Gamma\, t/2)$. 
Expressing the Kraus operators in this limit as, 
\begin{eqnarray*}
\label{akr}
K_0(t)&\approx& I-\frac{\Gamma\, t}{2}\, L_0,\ \ L_0=\left(\begin{array}{cccc} 
2 & 0 & 0 &0 \\ 
0 & 1 & 0 & 0 \\ 
0 & 0 & 1 & 0 \\ 
0 & 0 & 0 & 0 
\end{array} \right),   \nonumber \\
K_1(t)&\approx& \sqrt{\Gamma\, t}\, L_1, \ \ L_1=\left(\begin{array}{cccc} 
1 & 0 & 0 &0 \\ 
0 & 0 & 0 & 0 \\ 
0 & 0 & 1 & 0 \\ 
0 & 0 & 0 & 0
\end{array} \right), \nonumber \\ 
K_2(t)&\approx& \sqrt{\Gamma\, t}\, L_2,\ \ L_2=\left(\begin{array}{cccc} 
1 & 0 & 0 &0 \\ 
0 & 1 & 0 & 0 \\ 
0 & 0 & 0 & 0 \\ 
0 & 0 & 0 & 0 
\end{array} \right), \nonumber \\ 
  K_3(t)&\approx& \Gamma\, t\, L_3, \ L_3=\left(\begin{array}{cccc} 
1 & 0 & 0 &0 \\ 
0 & 0 & 0 & 0 \\ 
0 & 0 & 0 & 0 \\ 
0 & 0 & 0 & 0
\end{array} \right), 
\end{eqnarray*}   
We thus obtain the LGKS master equation describing the dynamics as in (\ref{Lindblad}): 
\begin{eqnarray}
\label{aLm}
\frac{d \rho_{AB}}{d\, t}&=& \Gamma \, {\cal L}_M \, \rho_{AB},\nonumber \\ 
&=& \Gamma (L_1\, \rho_{AB}\, L^\dag_1 + L_2\, \rho_{AB}\, L^\dag_2)-\frac{\Gamma}{2}\, (L_0\, \rho_{AB}+
\rho_{AB}\, L_0),\nonumber \\ 
 & & \ \ L_0=L_1^\dag L_1+L_2^\dag\, L_2. 
\end{eqnarray}

The fidelity function $F[\rho_{AB}(t),\rho_{AB}(t+\tau)]$ may be readily evaluated for the two qubit state (\ref{ye1rho}): 
\begin{widetext}
\begin{equation} 
F[\rho_{AB}(t),\rho_{AB}(t+\tau)]=\frac{1}{81}\, \left\{1+\sqrt{[4+\lambda\,\gamma^2(t)][4+\lambda\,\gamma^2(t+\tau)]}
+\sqrt{[4-\lambda\,\gamma^2(t)][4-\lambda\,\gamma^2(t+\tau)]}\right\}^2.
\end{equation}  
\end{widetext} 
The variation of fidelity as a function of dimensionless scaled time $\Gamma\, t$ is shown in Fig.~1. It may be seen that $F[\rho_{AB}(t),\rho_{AB}(t+\tau)]$ increases from its initial value $F[\rho_{AB}(0),\rho_{AB}(\tau)]$  and approaches unity asymptotically -- which is a typical Markovian behavior.

\medskip 

\noindent (b) Recently Yu an Eberly~\cite{YE2} considered a slight variation of the above  model leading to non-Markovian noise. The Kraus operators assoicated with the noisy evolution of two qubit system are obtained by replacing $\gamma(t) \rightarrow p(t)={\rm exp}[-f(t)], \ f(t)=\frac{\Gamma}{2}[t+\frac{1}{\gamma}(e^{-\gamma t}-1)],$  $\omega(t)\rightarrow q(t)=\sqrt{1-p^2(t)}.$ In this non-Markovian model, $\gamma$ denotes the environmental noise bandwidth and $\Gamma$  is the noise property assoicated with the qubit. In the limit $\gamma\rightarrow \infty,$ we get $f(t)\rightarrow \frac{\Gamma}{2}t$, and hence the Markovian dynamics is recovered.

In the short time limit viz., $\gamma\, t<<1,$ we have, $p(t)\approx 1-\frac{\Gamma\gamma\, t^2}{4}$ and $q(t)\approx t\,\sqrt{\frac{\Gamma\gamma}{2}}$ and clearly, this structure does not lead to the standard LGKS master equation and hence brings out the non-Markovian nature of the model. In the short time limit, the same operators as in (\ref{aLm}) appear --  but in the following form,
\begin{eqnarray} 
\rho_{AB}(t)-\rho_{AB}(0)&\approx& \frac{\Gamma\gamma\, t^2}{2}\, (L_1\, \rho_{AB}\, L^\dag_1 + L_2\, \rho_{AB}\, L^\dag_2)\nonumber \\ &&-\frac{\Gamma\gamma\, t^2}{4}\, (L_0\, \rho_{AB}+
\rho_{AB}\, L_0),\nonumber \\ 
{\rm or}\ \ \ \   \frac{d\rho_{AB}}{dt}&=&\frac{\gamma\,\Gamma\, t}{2}\, {\cal L}_M\, \rho(0).   
\end{eqnarray} 
 (Here,  ${\cal L}_M$ is the same as in (\ref{aLm})). In other words, one may recast the dynamical equation in this model as a non-Markovian master equation, with a linear time pre-factor in the LGKS master equation (\ref{Lindblad}).

\begin{figure}[h]
\includegraphics*[width=3in,keepaspectratio]{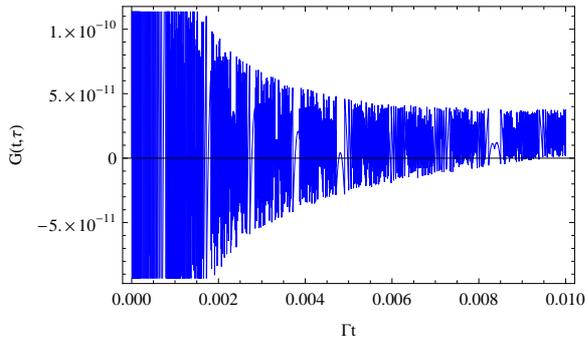}
\caption{(Color online). Fidelity difference $G(t,\tau)$  as a function of the dimensionless scaled time $\Gamma t$, in the  non-Markovian limit $\gamma=10^{-4}.$ (Here, we have chosen the parameters $\Gamma=1$ and $\tau=1$). Negative values of this function imply violation of the inequality (\ref{inequality}) and hence indicate non-Markovianity. All quantities are dimensionless.}
\end{figure}

Choosing a simple initial two-qubit state in the  $X$ form~\cite{YE2} 
\begin{eqnarray}
\label{ye1rho2}
\rho_{AB}(0)=\frac{1}{3}\, \left(\begin{array}{cccc} 
\alpha & 0 & 0 &0 \\ 
0 & 1 & 1 & 0 \\ 
0 & 1 & 1 & 0 \\ 
0 & 0 & 0 & 1-\alpha 
\end{array} \right),  
\end{eqnarray}
(where $\alpha$ denotes a real, positive parameter) the dynamics does preserve the $X$ structure, with the diagonal elements of the density matrix remaining unaltered and the off-diagonal elements acquiring a time dependence $[\rho_{AB}(t)]_{kl}=[\rho_{AB}(0)]_{kl}\, p^2(t)$. The fidelity associated with $\rho_{AB}(t)$ may  be readily evaluated to be,    
 \begin{widetext}
 \begin{equation} 
F[\rho_{AB}(t),\rho_{AB}(t+\tau)]=\frac{1}{9}\left\{1+\sqrt{[1+p^2(t)][1+p^2(t+\tau)]}+q(t)q(t+\tau)\right\}^2.
  \end{equation}
\end{widetext}

In Fig.~2 we have plotted the fidelity difference $G(t,\tau)$ (see Eq.(\ref{fd})) as a function of dimensionless scaled time $\Gamma t$, in the non-Markovian limit $\gamma <<1$. Negative fidelity difference reveal the non-Markovian feature.    

\begin{figure}
\includegraphics*[width=3in,keepaspectratio]{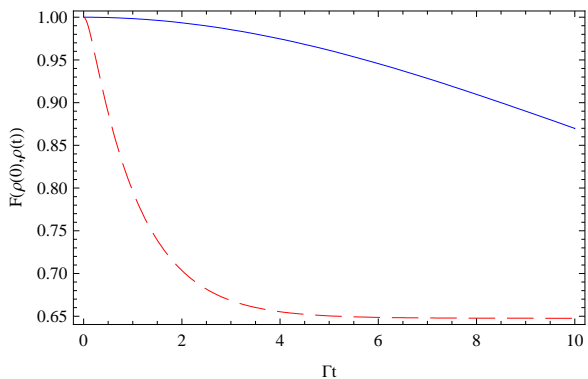}
\caption{(Color online). Fidelity $F[\rho(0),\rho(t)]$ of the  initial state $\rho(0)$ with dynamically density
matrix $\rho(t)$, as a function of dimensionless scaled time $\Gamma t$, in both Markovian (dashed curve; $\gamma=10$) and non-Markovian (solid curve; $\gamma=0.01$) limits; we have also chosen the parameter $\Gamma=1.$  It may be seen that the fidelity  larger in the non-Markovian case, when
compared to that in the corresponding Markovian case. All quantities are dimensionless.}
\end{figure}

We also find that the fidelity of the initial state with the dynamically evolved density matrix viz., $F[\rho(0),\rho(t)]$, has  larger value when $\gamma <<1$ (non-Markovian limit) compared to that in the limit $\gamma>>1$ (Markovian case) highlighting that the memory of the initial state is larger during non-Markovian dynamics.  This is depicted in Fig.~3, where we have plotted $F[\rho_{AB}(0),\rho_{AB}(t)]$   as a function of $\Gamma\,t$ both in the Markovian and non-Markovian limits.

\medskip 

(c)  Jaynes-Cummings model (JCM)~\cite{AKRR,AKRU} is a model of a two-level atom (qubit) interacting with a radiation field. 
This example, unlike the three models discussed above, starts with the Hamiltonian for the system for which the evolution operator can be constructed and we will examine here  the master equations for both the atom and photon subsystems. Incidentally, in this model, there is  sudden death of entanglement of the qubit with the radiation field~\cite{AKRR} -- a characteristic of non-Markovian evolution.  

 For simplicity of presentation, we consider here the resonant case where the qubit energy is equal to that of the radiation and the initial state of the atom is taken to be its excited state, $\rho_{A}(t=0)=\vert\uparrow\rangle\langle \uparrow\vert$. The initial state of the radiation is taken to be in a coherent state $\rho_{R}(t)=\vert\alpha\rangle\langle \alpha\vert,$   
$\vert\alpha\rangle=e^{-\frac{\vert\alpha\vert^2}{2}}\, \sum_{n=0}^{\infty}\frac{\alpha^n}{\sqrt{n!}}\vert n\rangle;$ $\vert\alpha\vert$ denoting the intensity of the radiation.  The Kraus representation for both the qubit and the radiation subsystems are explored in Ref.~\cite{AKRU}.

The dynamically evolved qubit density matrix is given by the mixed state~\cite{AKRR, AKRU}:
\begin{eqnarray} 
\label{atom} 
\rho_{A}(t)&=&\sum_{N=0}^{\infty}\, K_N(t)\rho_A(0)\,K^{\dag}_N(t),\nonumber \\ 
\label{atomKraus}
K_N(t)&=&W_{N\uparrow}(t)\, \vert\uparrow\rangle\langle\uparrow\vert\, +W_{N\downarrow}(t)\, \vert\downarrow\rangle\langle\uparrow\vert \\ 
\label{atomKrausele}
W_{N\uparrow}(t)&=&\cos(gt\sqrt{N+1})\, \langle N\vert\alpha\rangle,\nonumber \\ 
 W_{N\downarrow}(t)&=&-i\, \sin(gt\sqrt{N})\, \langle N-1\vert\alpha\rangle  \\
 \label{rhot} 
{\rm or} \ \ \rho_{A}(t)&=& \vert a\vert^2\, \rho(0)+\vert b\vert^2\, \sigma_-\rho(0)\, \sigma_{+}\nonumber \\ 
&& + i\, ab^*\, \rho(0)\, \sigma_{+}-ia^*b\, 
 \sigma_-\rho(0), 
\end{eqnarray} 	
Here,  $g$ denotes the interaction strength of the radiation with the qubit and  
\begin{eqnarray}
\label{par}
\vert a\vert^2&=&\sum_{N=0}^{\infty}\vert\langle N\vert\alpha\rangle\vert^2\, \cos^2(gt\sqrt{N+1}), \nonumber  \\
\vert b\vert^2&=&\sum_{N=0}^{\infty}\vert\langle N-1\vert\alpha\rangle\vert^2\, \sin^2(gt\sqrt{N}), \nonumber  \\
ab^*&=&\sum_{N=0}^{\infty}\, \cos(gt\sqrt{N+1})\sin(gt\sqrt{N}).  
 \end{eqnarray} 
 In the small time limit, the associated Kraus operators $K_N(t)$ of (\ref{atomKraus}) go as the square of time and as such, we do obtain a LGKS-type Master equation with a linear time prefactor -- as above in model (b) -- indicating a non-Markov feature.  

We now focus on a much simpler situation, where the initial radiation state is chosen to be the vacuum state $\rho_R(0)=\vert 0\rangle\langle 0\vert$,  to illustrate the non-Markovian behavior from the fidelity consideration.  The dynamically evolved qubit state then assumes the following simple form (obtained by substituting $\alpha=0$ in (\ref{rhot}), (\ref{par})) 
\begin{equation}
\label{simple} 
 \rho^{\uparrow}_A(t)=\cos^2(gt)\, \vert\uparrow\rangle\langle\uparrow\vert + \sin^2(gt)\, \vert\downarrow\rangle\langle\downarrow\vert  
\end{equation}   
in which situation the fidelity $F[\rho^{\uparrow}_A(t), \rho^{\uparrow}_A(t+\tau)]$ is readily found to be, 
\begin{eqnarray}
\label{fidatom}
 F[\rho^{\uparrow}_A(t), \rho^{\uparrow}_A(t+\tau)]&=&\left\{\vert \cos(gt)\cos[g(t+\tau)]\vert \right. \nonumber \\ 
 &&  \left. +\vert \sin(gt)\sin[g(t+\tau)]\vert\right\}^2
\end{eqnarray} 
The fidelity difference $G(t, \tau)$ (see Eq.~(\ref{fd}))  plotted in  Fig.~4, reveals negative fluctuations and hence is a clear manifestation of non-Markovian evolution. 
\begin{figure}
\includegraphics*[width=3in,keepaspectratio]{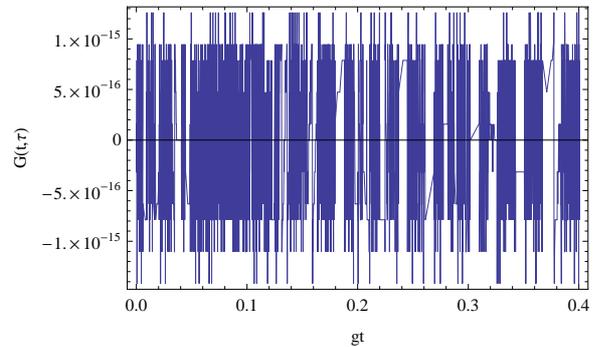}
\caption{(Color online). Fidelity difference $G(t, \tau)$ in  JCM with an initially excited atomic system, as a function of dimensionless scaled time $gt$. We have chosen the parameter $g\tau=10$. Negative fluctuations of the function $G(t, \tau)$ reveal non-Markovian behavior. All quantities are dimensionless.}
\end{figure}   

It would be interesting to explore the evolution of radiation subsystem as well in this model.  For simplicity we choose the same initial states of the atom (excited state)  and the radiation field (coherent state)  to obtain the dynamical state of the photon as, 
\begin{equation} 
\label{photon} 
\rho_R(t)=V_{\uparrow\uparrow}(t)\rho_R(0)V^\dag_{\uparrow\uparrow}(t)+V_{\downarrow\uparrow}(t)\rho_R(0)V^\dag_{\downarrow\uparrow}(t), 
\end{equation}
where the corresponding  Kraus operators are given by, 
\begin{eqnarray}
\label{phkr}
 V_{\uparrow\uparrow}(t)&=&\sum_{n=0}^{\infty} \cos(gt\sqrt{n+1})\, \vert n\rangle\langle n\vert\nonumber \\
 V_{\downarrow\uparrow}(t)&=&-i\sum_{n=0}^{\infty} \sin(gt\sqrt{n+1})\, \vert n+1\rangle\langle n\vert. \nonumber \\
\end{eqnarray}  
The small time behavior of the photon Kraus operators could be readily recognized as, 
\begin{eqnarray}
 V_{\uparrow\uparrow}(t)&\approx& I_R-\frac{1}{2} (gt)^2\, L_0 \nonumber \\
 V_{\downarrow\uparrow}(t)&\approx& -i\, gt \, L_1,  \nonumber \\
\end{eqnarray} 
where the Lindblad operators $L_0, L_1$ are related to the photon creation and annihilation operators $a^\dag, \ a$ as follows: 
\begin{eqnarray*}
 L_0&=& a\, a^\dag =\sum_{n=0}^{\infty}\, (n+1) \vert n\rangle\langle n\vert \\
 L_1&=&a^\dag=\sum_{n=0}^{\infty}\, \sqrt{n+1}\, \vert n+1\rangle\langle n\vert. \\
\end{eqnarray*}
Just as in the case of  qubits, we  get the LGKS type master equation with a linear time prefactor -- indicating a non-Markovian behavior. 
\medskip 

\noindent (d) Chruscinski and Kossakowski~\cite{CK2}  presented an interesting model to elucidate non-Markovian quantum dynamics described either by non-local master equation or by a local time formulation. Here, the master equation governing the dynamics of a single qubit system given by, 
\begin{equation} 
\label{master} 
\frac{d\rho}{dt}=\int_{t_0}^{t}{\cal K}(t-u)\rho(u)du,
\end{equation}  
consists of a  {\it time independent} memory kernel ${\cal K}(t)= \frac{1}{2}{\cal L}_0$ where ${\cal L}_0$ is a pure dephasing generator, 
\begin{equation}
{\cal L}_0\rho=\sigma_z\rho\sigma_z-\rho. 
\end{equation}  
(Here $\sigma_z$ denotes the $z$ component of Pauli spin operator of the qubit.) 

In an equivalent  approach, the completely positive, trace preserving map $\Lambda(t,t_0)$ characterizing the dynamics $\rho(t)=\Lambda(t,t_0)\rho(t_0)$ satisfies a local in time equation,   
\begin{equation} 
\label{local} 
\frac{d\Lambda(t,t_0)}{dt}={\cal L}(t-t_0)\Lambda(t,t_0),
\end{equation} 
in terms of a highly singular generator 
\begin{equation}
{\cal L}(t-t_0)=\tanh(t-t_0)\, {\cal L}_0. 
\end{equation}  
Despite the fact that the local in time dynamics involves a singular generator, the dynamical map has a regular solution given by~\cite{CK2}, 
\begin{equation}
\Lambda(t,t_0)=\frac{1}{2}[1+\cos(t-t_0)]\, I + \frac{1}{2}[1-\cos(t-t_0)]\, ({\cal L}_0+I) 
\end{equation}  
and the evolved qubit density matrix is therefore obtained as, 
\begin{equation}
\rho(t)=\Lambda(t,0)\rho(0)=\left(\begin{array}{cc} \rho_{11}(0) & \rho_{12}(0)\cos t \\ 
\rho_{12}^*(0)\cos t & \rho_{22}(0)
\end{array} \right), 
\end{equation}
exhibiting oscillations in qubit coherence. 

The above dynamics may also be characterized in terms of a two element Kraus operator set
\begin{eqnarray}
K_0(t)=\cos(t/2)\, I= \left(\begin{array}{cc} \cos(t/2) & 0 \\ 
0 & \cos(t/2)
\end{array} \right)\nonumber \\ 
K_1(t)=\sin(t/2)\, \sigma_z= \left(\begin{array}{cc} \sin(t/2) & 0 \\ 
0 & -\sin(t/2)
\end{array} \right)
\end{eqnarray}   
leading to the dynamical evolution $\rho(t)=\sum_{i=0,1}K_i(t)\rho(0)K_i^\dag(0)$.
Evidently the small time form of the Kraus operators ($K_0(t)\approx I(1-t^2/2)$ and $K_1(t)\approx \frac{t}{2}\, \sigma_z$) lead to a master equation of the LGKS form -- with a linear time pre-factor.  

\begin{figure}
\includegraphics*[width=3in,keepaspectratio]{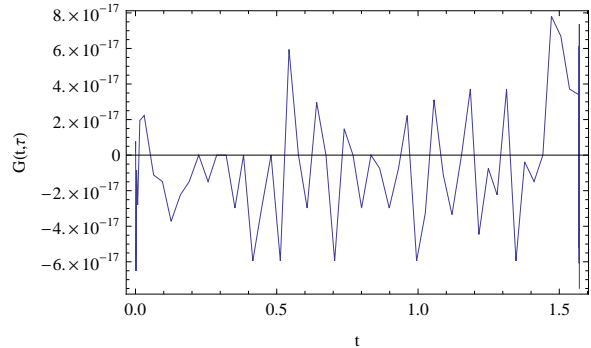}
\caption{(Color online). Fidelity difference $G(t, \tau)$ corresponding to the dynamical state (\ref{qu}), as a function of (dimensionless) time $t$; here, $\tau=\pi/6$. Negative values of $G(t,\tau)$ point towards non-Markovian behavior. All quantities are dimensionless.}
\end{figure}   

Further, considering the initial qubit state to be a pure state with $\rho_{11}(0)=\rho_{22}(0)=\rho_{12}(0)=\frac{1}{2}$, we obtain the evolved system as, 
\begin{equation}
\label{qu}
\rho(t)=\frac{1}{2}\left(\begin{array}{cc} 1 & \cos(t) \\ 
\cos(t) & 1
\end{array} \right),
\end{equation}    
We obtain the fidelity $F[\rho(t),\rho(t+\tau)]$ as, 
\begin{eqnarray}
F[\rho(t),\rho(t+\tau)]&=&\frac{1}{2}\left(1+\cos t \cos(t+\tau)\right. \nonumber \\
 && \ \ \left. +\vert \sin t \sin(t+\tau) \vert  \right).
\end{eqnarray}
We have plotted the fidelity difference $G(t,\tau)$ in Fig.~5. The negative values assumed by the fidelity difference $G(t,\tau)$ (see Fig.~5) point towards the violation of the inequality (\ref{inequality}) -- which highlights the non-Markovian incarnation in this model.

\section{Conclusions} 
From the Kraus representation of the dynamical evolution and  fidelity as a 
measure of determining the propensity of the initial state in the time evolved state, we have elucidated the manifestation of Markovian or non-Makovian incarnations. We have also proposed fidelity difference to capture the essence of non-Markovianity.  With the help of some examples, we have explored the nature of small time behavior of the dynamical Kraus form of quantum dynamics, which covers both  Markovian and non-Markovian processes (in the conventional sense)  depending on the form of interaction of the system with its environment as well as its initial state. We have shown that in the density matrix evolution governed by non-Markovian dynamics, the fidelity difference fluctuates between positive and negative values -- a clear signature of non-Markovianity. Moreover,  the memory of the initial state  carried by the dynamically evolving state -- characterized in terms of the fidelity  --  is shown to be  larger in the non-Markovian limit compared to that in the Markovian case. These two features viz., the small time behavior of Kraus operators and the fidelity, together confirm the Markov or non-Markov behavior in a consistent way.


\begin{thebibliography}{0}
\bibitem{note} Note that in the case of Markovian dynamics, the inequality (\ref{inequality}) is obeyed  when $t$ and $\tau$ are interchanged.
\bibitem{note2} The trace distance $D(\rho_1,\rho_2)$ is bounded from above and below in terms of the fidelity $F(\rho_1,\rho_2)$ as~\cite{NC}, $1~-~\sqrt{F(\rho_1,\rho_2)}\leq D(\rho_1,\rho_2)\leq \sqrt{1-F(\rho_1,\rho_2)}.$ In the limit when fidelity (overlap between two states) approaches the value unity, the trace distance vanishes.
\end{thebibliography}
\end{document}